\magnification\magstep1
\font\BBig=cmr10 scaled\magstep2


\def\title{
{\bf\BBig
\centerline{Topology}
\bigskip
\centerline{of}
\bigskip
\centerline{non-topological Chern-Simons vortices}
}
} 


\def\author{
\centerline{
P.~A.~HORV\'ATHY  
}
\bigskip
\centerline{
Laboratoire de Math\'ematiques et de Physique Th\'eorique}
\medskip
\centerline{Universit\'e de Tours}
\medskip
\centerline{Parc de Grandmont,
F--37200 TOURS (France)
\foot{e-mail: horvathy@univ-tours.fr}
}
}

\def\runningauthor{
Horv\'athy 
}

\def\runningtitle{Topology of non-topological vortices
}


\voffset = 1cm 
\baselineskip = 14pt 

\headline ={
\ifnum\pageno=1\hfill
\else\ifodd\pageno\hfil\tenit\runningtitle\hfil\tenrm\folio
\else\tenrm\folio\hfil\tenit\runningauthor\hfil
\fi
\fi}

\nopagenumbers
\footline={\hfil} 


\def\and{\qquad\hbox{and}\qquad}

\def\kikezd{\parag\underbar} 
\def\IC{{\bf C}}
\def\IR{{\bf R}}
\def\IS{{\bf S}}
\def\IZ{{\bf Z}}
\def\Sz{{\bf S}_{z}}
\def\Sw{{\bf S}_{w}}
\def\Tr{{\rm Tr\,}}

\def\smallover#1/#2{\hbox{$\textstyle{#1\over#2}$}}
\def\2{{\smallover 1/2}}
\def\parag{\hfil\break} 
\def\p{\partial}

\def\smallcirc{{\raise 0.5pt \hbox{$\scriptstyle\circ$}}}

\def\sk{({\rm sign\ }\kappa)}

\newcount\ch 
\newcount\eq 
\newcount\foo 
\newcount\ref 

\def\chapter#1{
\parag\eq = 1\advance\ch by 1{\bf\the\ch.\enskip#1}
}

\def\equation{
\leqno(\the\eq)\global\advance\eq by 1
}

\def\foot#1{
\footnote{($^{\the\foo}$)}{#1}\advance\foo by 1
} 

\def\reference{
\parag [\number\ref]\ \advance\ref by 1
}

\ch = 0 
\eq=1
\foo = 1 
\ref = 1 


\title
\vskip10mm
\author
\vskip.25in

\parag
{\bf Abstract.}
{\it  The quantized magnetic flux
$\Phi=-4\pi\,N\sk,\, N=0,\pm1,\dots$ of non-topological
vortices in the non-relativistic Chern-Simons theory
is related to the topological degree of the $S^2\to S^2$ mapping
defined by lifting the problem to the Riemann spheres.
Regular solutions with finite degree only arise for 
rational functions, 
whose topological degree, $N$, is the commun number of their zeros
and poles on the Riemann sphere, also
called their algebraic order.}

\vskip15mm
\noindent
(\the\day/\the\month/\the\year)
\vskip15mm
\medskip\noindent
\vskip15mm 

\vskip5mm

\vfill\eject

A striking feature of non-relativistic Chern-Simons vortices
[1] is the quantization of their magnetic flux,
$$
{\rm vortex\ flux}=\int_{\IR^2}\!B\,d^2\vec{x}=-4\pi\, N\sk,
\qquad
N=0,\pm1,\dots
\equation
$$

In gauge theories, similar properties are usually
related to {\it topology}. 
Yang--Mills--Higgs 
theories\foot{We assume, for simplicity, that the gauge group is {\rm 
SU}$(2)$.},
for example, support finite--energy monopole solutions [2]. 
For large $r$, the Higgs field is required to
satisfy $\Tr(\Phi)^2\-=1$ rather then to vanish.
This asymptotic condition allows us to view
$\Phi$ as a mapping from the two-sphere at infinity, 
$\IS\equiv\IS^2_{\infty}$,
into the vacuum manifold $\Tr(\Phi)^2\-=1$, which is again a two-sphere.
Then the magnetic charge of the monopole is expressed [3] as
$$
{\rm Monopole\ charge}={1\over8\pi}\int_{\IS}
\epsilon_{ijk}\epsilon_{abc}\Phi^a\p_{j}\Phi^b\p_{k}\Phi^c\,d^2S.
\equation
$$

The r. h. s. here is a topological invariant characterized by an 
integer.
It is in fact the homotopy class (or ``winding number'') of $\Phi$,
$[\Phi]\in\pi_{2}(\IS^2)\simeq\IZ$. 

Non-relativistic Chern-Simons vortices are,
however, {\it non-topological}~: for finite--energy solutions
the particle density, $\varrho$, 
tends to zero when spatial infinity is approached.
Then where does flux quantization, Eq. (1), come from~?

To answer this question, let us remember that, owing to the 
``field-current identity'' which relates the
magnetic field to the particle density, $\kappa B=-\varrho$, 
the magnetic flux of the vortex is 
$$
{\rm vortex\ flux}=-{1\over\kappa}\int_{\IR^2}\!\varrho\,d^2\vec{x}.
\equation
$$

Here, apart of the zeros, $\varrho$ is a solution of the Liouville 
equation
$
\bigtriangleup\log\varrho=-(2/\vert\kappa\vert)\varrho.
$
Now the general solution
of the Liouville equation is expressed using an arbitrary analytic 
function
$f(z)$, 
$$
\varrho=
{4\vert\kappa\vert}{\vert f'\vert^2\over(1+\vert f\vert^2)^2}, 
\equation
$$
so that 
$$
-{\sk\over4\pi}\,\big({\rm vortex\ flux}\big)=
{1\over4\pi}
\int_{\IR^2}\!{4\vert f'\vert^2\over(1+\vert f\vert^2)^2} 
\,dxdy.
\equation
$$

The point is that the r. h. s. has the {\it same} topological 
interpretation  
as (2) for mono\-poles.
The analytic function $f$ can in fact be viewed as
 mapping  the $z$-plane into the $w$-plane.
Compactifying these planes into Riemann spheres
$\Sz\simeq\IC\cup\{\infty\}$ and 
$\Sw\simeq\IC\cup\{\infty\}$,  
$z$ and $w$ become stereographic coordinates.
If $f$ admits a well--defined (finite or infinite) limit as 
$r\to\infty$, it can be extended into 
a mapping from $\IS\equiv\Sz$ 
into $\Sw$ we still denote by $f$.
The extended mapping has a winding number, 
 given in (2). This latter is in fact the integral of 
the pull-back by $f$ of $\Omega$,
the surface form of the unit sphere $\Sw$, 
$$
{\rm charge}={1\over4\pi}\int_{\IS}f^*\Omega.
\equation
$$
Now, in complex coordinate $w$,
the surface element of $\Sw$ is
$$
\Omega=
2i{dw\wedge d\overline{w}\over(1+w\overline{w})^2},
\equation
$$
so that setting $w=f(z)$ in Eq. (6) yields precisely the r. h. s. of
(5), as stated.
\goodbreak

In a previous paper [4] we proved that the regularity of finite--flux 
 solutions excludes essential singularities in the finite domain as 
 well as at infinity. 
An analytic function whose only singularities in the extended plane 
are poles is known to be a {\it rational function},
$$
f(z)={a_{m}z^m+\dots+a_{1}z+a_{0}
\over
b_{n}z^n+\dots+b_{1}z+b_{0}}.
\equation
$$
 [5].
But a rational function 
has always a limit at infinity, namely $\infty\in\Sw$ if $m>n$,
$0\in\Sw$ if $m<n$ and $a_{m}/b_{m}\neq0,\,\infty$ if $m=n$.
In the first case, $\infty\in\Sz$ is a pole of order $m-n$
while in the second it is a zero of order $n-m$.
Since the only other poles and zeros are the $n$ zeros of the denominator
and the $m$ zeros of the numerator,  respectively, we see that
the extended map $f~:\Sz\to\Sw$ has the same number 
of  zeros and poles,
namely  $N=$ {\rm max}$(m,n)$. 
This commun number is called the
{\it algebraic order} of the rational map $f$ [5].

The magnetic charge (2) has another useful desciption [3].
The r. h. s. of (2) is in fact equal to the
{\it topological} (or Brouwer) {\it degree}
of the field $\Phi$. Chosing coordinates
on the spheres, the mapping $\Phi$ can be characterized
by $\Phi^{a}(\xi)$, $\xi=(\xi^{b})\, a,\, b=1,2$.
A point $\Phi_{0}$ in the image--sphere is
a regular point if
${\rm det}\big(\p\Phi^{a}/\p\xi^{b})\neq0$).
Then the topological degree is the number of 
pre-images of a regular point $\Phi_0$
$$
{\rm degree}=\sum_{\xi\in\Phi_{0}}{\rm sign\ det}\Big(
{\p\Phi^{a}\over\p\xi^{b}}\Big).
\equation
$$
 
Eq. (9) counts the number of times the image--two--sphere is covered while
the domain--two--sphere is covered once. 
Then, assuming that $\infty$ and $0$ are regular points,
 the topological degree of the rational function (8) is  simply the 
number of zeros or, equivalently, the number of poles.
Thus, we have proved 

\kikezd{Theorem1}.
{\it The magnetic flux of the vortex solution associated with the 
rational function $f$ is}
(1), {\it where $N$ is the 
algebraic order of $f$}.

\vskip2mm
The same result was obtained in Ref. [4]  by a rather laborious
direct calculation. 

Let us also remark that the vortex number could 
also be identified with the first Chern class of an
${\rm U}(1)$ principal bundle over the two--sphere.

The correspondence between analytic functions $f$ and solutions 
$\varrho$ of the Liouville equation is not one--to--one, though.
Let  $R$ denote an arbitrary rotation of the image--two--sphere
$\Sw$. Then
$f$ and $R\smallcirc f$ yield  the same solutions.
This is seen by noting that the particle density 
is in fact the dual on  
the two-sphere $\IS\equiv\Sz$ of the two-form $f^*\Omega$,
$\varrho=\star\big(f^*\Omega\big)$. But
$$
\big(R\smallcirc f\big)^*\Omega
=f^*\big(R^*\Omega\big)=f^*\Omega,
$$
because the surface--form $\Omega$ is symmetric w. r. t. rotations,
$R^*\Omega=\Omega$.
There is hence no loss of generality in requiring that 
$$
f(\infty)=0,
\equation
$$
since this can always be achieved by a rotation of $\Sw$.
 A rational function which satisfies the condition (10) can be written as
$$
f(z)={P(z)\over Q(z)},
\qquad
{\rm deg}\,P < {\rm deg}\,Q
\equation
$$
whose order is 
$N={\rm deg}\,Q$. Such an $f$ depends on the
$2N$ complex parameters, namely the coefficients of the polynomials
$P(z)$ and $Q(z)$. 
Thus, we have proved

\kikezd{Theorem2}.
{\it For fixed  magnetic flux, the solution solution depends on $4N$ 
real parameters}.

\vskip2mm
Requiring $f(\infty)=\infty$ instead of
(10), the role of poles 
and zeros would be interchanged, without changing the conclusion.

Theorem2 is again consistent with previous results,
[4], [6]. The particular form (11) was reached in Ref. 
[4] by exploiting the invariance of the solution by
$f\to (1/f)$ when $m >n$ in (8), 
and by reducing the case $m=n$
 to (11) by a suitable redefinition. 
These transformations correspond simply to ${\rm O}(3)$ 
transformations
which interchange poles and zeros or rotate a point
$\neq0,\, \infty$ into the origin, respectively.

Let us note, in conclusion, that our result also applies to 
the ${\rm O}(3)$ sigma model which describes a
two-dimensional ferromagnet [7]. The scalar field $\Phi$ is here 
a mapping from the 
plane into a two-sphere $\IS^2$. For large distances, all 
magnets are required to be aligned so that
 $\Phi(\infty)=\Phi_{0}\in\IS^2$. This 
allows to extend $\Phi$ into a mapping between two-spheres.
 The self-dual solutions found by
Belavin and Polyakov  [7] are  associated with analytic 
functions on the complex plane with identical properties as here
above.

\goodbreak
\kikezd{Acknowledgement}. 
It is a pleasure to thank N. Manton and C. Duval for
illuminating discussions. 

\goodbreak
\vskip5mm

\centerline{\bf\BBig References}

\reference
R.~Jackiw and S-Y.~Pi,
{\sl Phys. Rev. Lett}. {\bf 64}, 2969 (1990);
{\sl Phys. Rev}. {\bf D42}, 3500 (1990).
For review see
R. Jackiw and S-Y. Pi, {\sl Prog. Theor. Phys. Suppl}. {\bf 107}, 1 
(1992),
or
G. Dunne, {\sl Self-Dual Chern-Simons Theories}. 
Springer Lecture Notes in Physics. New Series: Monograph 36. (1995).

\reference
For a review on monopoles see, for instance, 
P.~Goddard and D. Olive, {\it Monopoles in gauge theory},
{\sl Rep. Prog. Phys}. {\bf 41}, 1357-1437 (1978).

\reference
J.~Arafune, P~G.~O. Freund, and C. J. Goebel,
{\sl Journ. Math. Phys}. {\bf 16}, 433 (1975).
The topological facts used in this paper are taken from
J.~W.~Milnor,
{\sl Topology from the Differentiable Viewpoint},
Univ. of Virginia Press, Charlottesville (1965).

\reference
P.~A.~Horv\'athy and J.-C.~Y\'era,
{\sl Lett. Math. Phys}. {\bf 46}, 111-120 (1998).

\reference
L.~V.~Ahlfors, {\sl Complex analysis},
McGraw--Hill (1979). See also E.~T.~Whittaker and G. N. Watson,
{\sl A course of modern analysis}.
Fourth edition. Cambridge University Press.
Reprinted in 1992.

\reference 
S. K. Kim, K. S. Soh, and J. H. Yee,
{\sl Phys. Rev}. {\bf D42}, 4139 (1990).

\reference
A.~A.~Belavin, A.~M.~Polyakov, A.~S.~Schwartz and Yu.~S.~Tyupkin,
{\sl Phys. Lett.} {\bf 59B}, 85 (1975).
The hypersphere description was proposed by
R.~Jackiw and C.~Rebbi, {\sl Phys. Rev}. {\bf D14}, 517 (1976).
\goodbreak

\reference
A.~A.~Belavin and A.~M.~Polyakov, 
{\sl JETP. Lett.} {\bf 22}, 245 (1975).

\bye